\documentclass[%
 reprint,
superscriptaddress,
nofootinbib,
 amsmath,amssymb,
 aps,
floatfix
]{revtex4-2}

\usepackage{graphicx}
\usepackage{dcolumn}
\usepackage{bm}
\usepackage{hyperref}

\usepackage{siunitx}
\usepackage{hyperref}
\usepackage{ragged2e}
\usepackage{float}

\usepackage{xcolor}
\usepackage{soul}
\usepackage[normalem]{ulem}
\usepackage{subcaption}
\usepackage{chngcntr}
\usepackage{mwe}

\newcommand{\angstrom}{\textup{\AA}}

\newcommand{\Koa}{$\left|\langle K_{out} \rangle\right|$}

\newcommand\LSdeleted{\bgroup\markoverwith{\textcolor{blue}{\rule[0.5ex]{2pt}{0.4pt}}}\ULon}

\newcommand{\NAdeleted}[1]{\bgroup\setstcolor{brown}\st{#1}\egroup}

\begin{document}

\preprint{APS/123-QED}

\title{Blue Shifts in Helium-Surface Bound-State Resonances and Quantum Effects in Knudsen Scattering}

\author{Luke Staszewski}%
\affiliation{Former Address: Cavendish Laboratory, University of Cambridge, 19 J J Thomson Avenue, Cambridge CB3 0HE, United Kingdom}

\author{Nadav Avidor}%
\email{na364@cantab.ac.uk}
\affiliation{Former Address: Cavendish Laboratory, University of Cambridge, 19 J J Thomson Avenue, Cambridge CB3 0HE, United Kingdom}

\date{\today}

\begin{abstract}
The scattering of gas from surfaces underpins technologies in fields such as gas permeation, heterogeneous catalysis and chemical vapour deposition. The effect of surface defects on the scattering is key in such technologies, but is still poorly understood. It is known empirically that unordered surfaces result-in Knudsen, random-angle, scattering, with the effect thought to be classical. We here demonstrate the transition from quantum mechanical diffraction to an apparent Knudsen scattering, and show that quantum bound-state resonances can greatly affect this transition. Further, we find that randomly distributed defects induce a blue-shift in the bound-state energies. We explore this phenomena, which can lay the basis for helium based quantum metrology of defects in 2D materials and material surfaces.

\end{abstract}

\maketitle

The scattering of atoms from a solid surface is a fundamental process in gas-solid interactions, including in fields such as gas-surface permeation, chemical vapour deposition and heterogeneous catalysis \cite{catalyst,feres2004knudsen,keerthi2018ballistic}, to name a few. The scattering from ordered surfaces is diffractive and is well understood from quantum mechanics (QM). However, disordered surfaces present diffuse scattering, which in its extreme has cosine, random angle, scattering as described empirically by Knudsen \cite{knudsen1967cosine,comsa1968angular,wenaas1971equilibrium}. Despite applications in fields such as previously mentioned, the transition between diffractive to Knudsen scattering is poorly understood \cite{o1971atomic}. We here find that quantum mechanical effects play an important role in Knudsen scattering. Furthermore, we show that even when restricting the scattering to elastic only, the empirical phenomena (Knudsen scattering) can be obtained. Our results, as discussed below, hint on a mechanism for an apparent thermalisation of the helium-surface eigenstates \cite{deutsch, srednicki}.

Knudsen scattering is a result of a complete "loss of memory" with respect to the incident particle directionality. One straight forward mechanism which can explain this, is the adsorption of particles on the surface, which results "loss of memory" prior to desorption. However, for inert particles which do not adsorb, like helium atoms, this explanation does not hold. Previously, loss of directionality has been explained also by macroscopic effects such as scattering from porous surfaces \cite{feres2004knudsen}. The pores trap the gas particle, which scatters in the pore, looses memory of its original direction, and eventually scatters out of the pore and contribute to a cosine-scattering distribution.

More recently, a classical description of the surface at the nano-scale has emerged to describe Knudsen scattering. For example, Knudsen scattering was demonstrated by considering the particle-surface interaction from a more realistic point of view, by using an extended "washboard model" \cite{liang2013performance,liang2019accurate,mateljevic2009accommodation}. In this model, the particle-surface interaction potential is composed of a repulsive surface (hard-wall), and an attractive region which extends above that hard-wall. Scattering is then realised using molecular dynamics (MD) simulations, and the results serve as input to higher level models. In all such modeling, including MD, the scattering was assumed classical.

Recent experiments in scanning helium microscopy have demonstrated that, similar to more traditional diffraction experiments, while well ordered surfaces result-in the expected diffractive scattering \cite{bergin2020observation}, Knudsen scattering better explains imaging of rough surfaces \cite{lambrick2018ray,lambrick2020multiple}. These results, in addition to further experimental evidence for both diffuse and diffractive scattering \cite{keerthi2018ballistic,thiruraman2020gas}, suggest that QM effects may be important for Knudsen scattering.

In this paper we find that QM effects indeed have an important role in scattering appearing as Knudsen-like. By using close coupled equations to solve the time-independent Schr\"odinger equation, we simulate the scattering of helium atoms from various static surfaces. We demonstrate the transition from diffraction towards scattering with cosine distribution, and discuss the role of certain aspects of the gas-surface potential in resulting Knudsen scattering. Furthermore, we explore helium-surface bound-state resonances (BSR), which is an important aspect of the scattering,  and find that disorder induces an increase in the bound state energies (blue shift).

The scattering calculations were performed using the program "Multiscat", which solves the time-independent Schr\"odinger equation using close-coupling methods \cite{wolken1973theoretical,sanz2007selective}. Multiscat considers the helium wave-function in terms of a basis comprised of Lobatto shape functions, and solves the eigenvalue problem iteratively using a variational principle \cite{manolopoulos1990iterative,riley2008analysis}.

\begin{figure}
\includegraphics[width=0.4\textwidth]{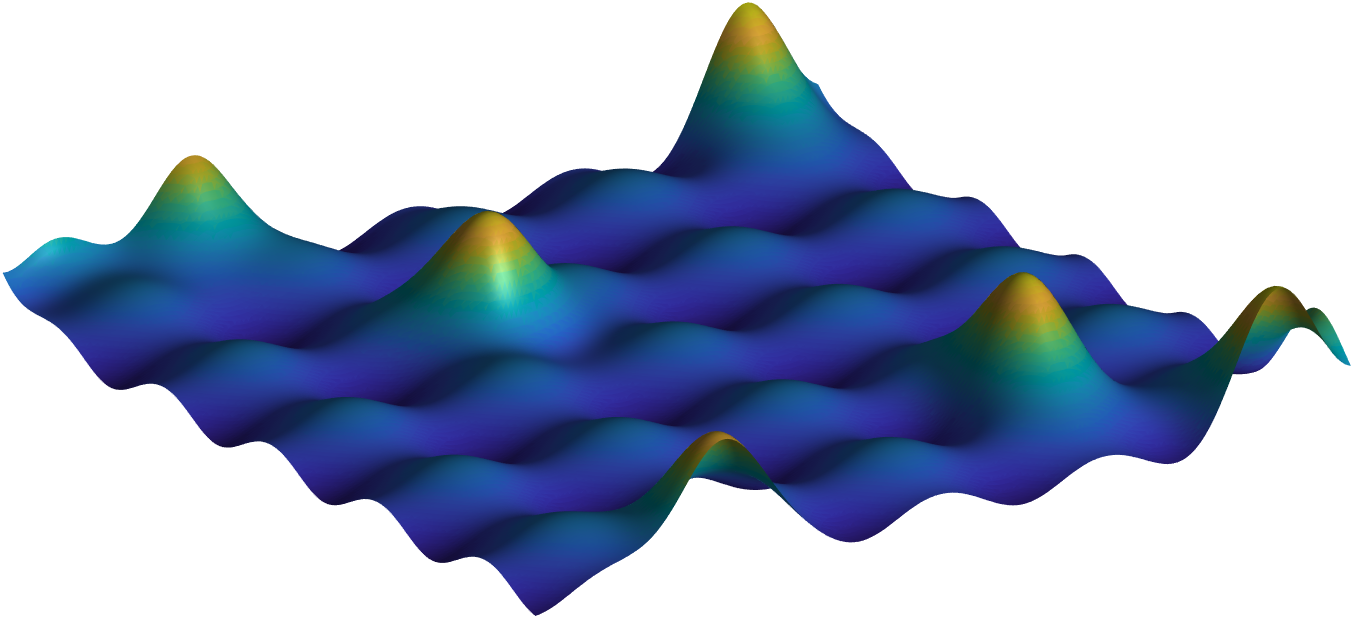}
\caption{\label{Knudsen_potential} Equipotential surface plot with 5 Gaussian shaped adatoms placed on a 5X5 supercell with lattice parameter of $2.84\angstrom$}
\end{figure}

Close-coupling methods were briefly used in the past for considering scattering from disordered surfaces \cite{whaley1994time}, however not in the context of Knudsen scattering. The close coupled formalism allows solving the scattering from realistic He-surface interaction potentials. We took the basis for our potential to be a corrugated Morse potential, of the form
\begin{equation}
    V(x,y,z) = D(Q(x,y) \cdot e^{-2 \alpha (z - z_0)} - 2e^{- \alpha (z - z_0)})
    \label{eqn:Vxyz}
\end{equation}
where $D [meV]$ is the potential depth, $\alpha [\AA^{-1}]$ is the stiffness parameter and $z_0$ is an arbitrary offset for the surface. The corrugation function, $Q(x,y)$ describes how the potential varies laterally along the surface. For our current investigation, we explored a potential based on previous numerical investigations of the He-LiF(100) interaction potential \cite{manolopoulos1990iterative} and take just the first few Fourier components to describe the corrugation:
\begin{equation}
Q(x,y)= 1+ \beta (cos( \frac{2 \pi x}{a})+cos( \frac{2 \pi y}{a}))
\label{eqn:Qxy}
\end{equation}.
The lattice constant, $a$, is set to $2.84\angstrom$ and the dimensionless factor, $\beta$, describing the strength of the corrugation is set to $0.04$ \cite{manolopoulos1990iterative}. On top of the LiF corrugation we then added randomly distributed adatoms. Adatoms were emulated using bivariate Gaussian modulations, to create local shifts in $z_0$ of equation \ref{eqn:Vxyz}. Such approximation of the adatoms neglects the local distortion of the attractive part of the potential, hence enable a more restrictive study of the effect of disorder on the scattering. An example potential is shown in Fig \ref{Knudsen_potential}.

\begin{figure}
\includegraphics[width=0.5\textwidth]{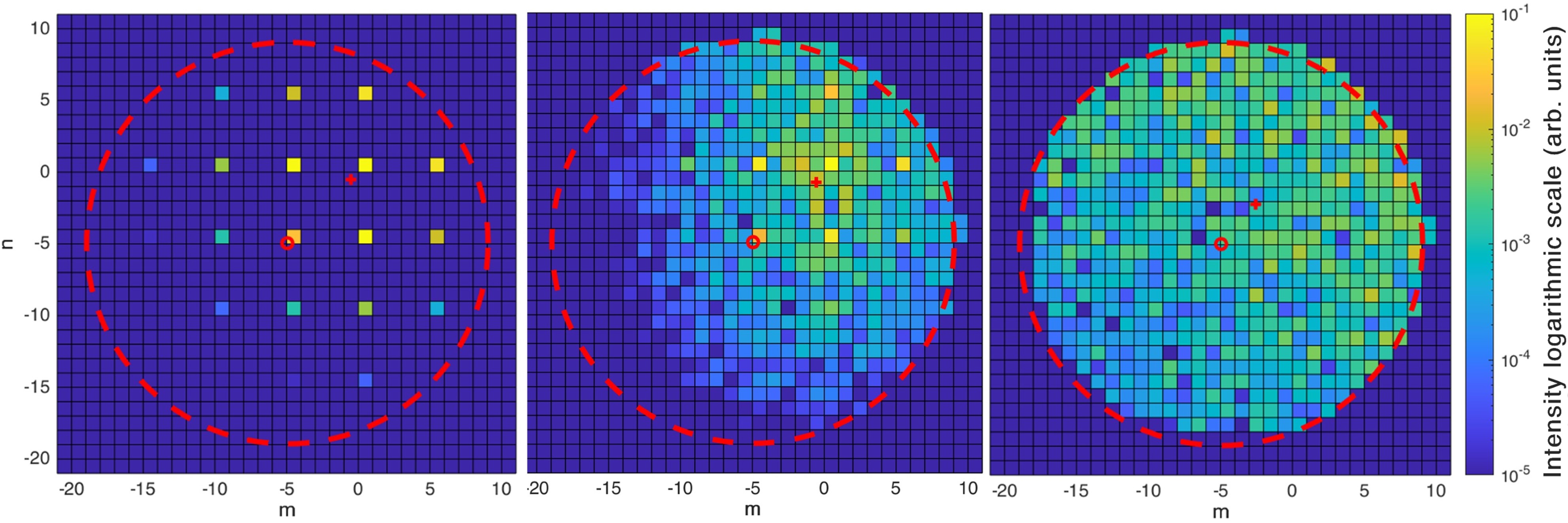}
\caption{\label{k_space_clean_and_dis} Shift from diffraction pattern for a clean surface (left) to diffuse scattering in accordance with Knudsen Scattering (right) with disorder parameters, $\theta$ (the number density of adsorbates) of 0, 0.2 and 3.2 respectively and an incident beam angle of $30^{o}$. Intensity indicated by colour is plotted against the index of the scattered wave. (0,0) corresponds to the specular peak. The disorder parameter corresponds to the concentration of adatoms randomly distributed within a supercell. The dotted red circle indicates an outgoing wave vector at $90^{o}$ to surface normal. This region should be filled with equal intensity for ideal Knudsen scattering. The red cross indicates the average of the outgoing momentum parallel to the surface and will be over the specular peak for highly reflective scattering and over the centre circle for ideal diffuse scattering.}
\end{figure}

We have investigated randomly distributed adatoms with and without preferential sites of adsorption. Following the construction of He-surface potentials, we have calculated the scattering. Representative scattering matrices for a defect-free (ordered) surface and a surface with randomly distributed defects are shown in Fig \ref{k_space_clean_and_dis}.

The (cosine) law of Knudsen states that the normalised intensity scattered into a solid angle, $d\Omega$, is given by: $dI(\gamma,\phi) = \frac{1}{\pi}cos(\gamma)d\Omega$. The equivalence of this law for the outputs of a scattering matrix is that each open diffraction channel should have equal intensities \cite{bergin2020observation}. This is equivalent to a maximum entropy principle for the scattered intensities. We therefore characterise the diffuse scattering by the Shannon entropy of the scattering matrix. The Shannon entropy is expressed as:
\begin{equation}
    S = -\frac{1}{\ln\left(\mathcal{N}\right)}\sum_{m,n}I_{m,n}\ln\left(I_{m,n}\right)
    \label{eqn:Shannon}
\end{equation}
where $\mathcal{N}$ is the total number of open channels and $I_{m,n}$ is the intensity in a given diffraction channel. This normalisation is such that the entropy is maximally unity.
The Shannon entropy does not give indication to the 'rate' in which the scattering tends to Knudsen, since even if the intensity is broadened around the specular condition, $S$ will increase to an extent. Therefore, in addition to $S$ we consider the magnitude of the average momentum parallel to the surface after scattering, \Koa, weighted by the intensity of the diffraction channels. \Koa will tend towards zero for diffuse scattering, where there is no preferential direction to scatter into. We note, however, that \Koa will also decrease when the intensity is shifted from the specular condition to some diffraction channels (for example due to coupling with bound-state resonances), which is not the same as 'tendency' towards Knudsen scattering. Note that \Koa takes on a non-zero value for a 'clean' surface as the incoming polar angle is $30^{o}$. In summary, only a combination of an increasing $S$, with a decreasing \Koa, can inform on tendency towards Knudsen scattering.

The effect of tendency towards Knudsen scattering upon introduction of partial disorder can be seen in Figure \ref{k_space_clean_and_dis}; the average momentum (parallel to the surface) tends towards zero as a function of disorder. We note that the convergence is not complete, and that closer convergence would have required a He-surface potential of much larger super-cell. However, accounting for such a super-cell with the required degree of corrugation was beyond the available computing resources.

\begin{figure}
\includegraphics[width=0.45\textwidth]{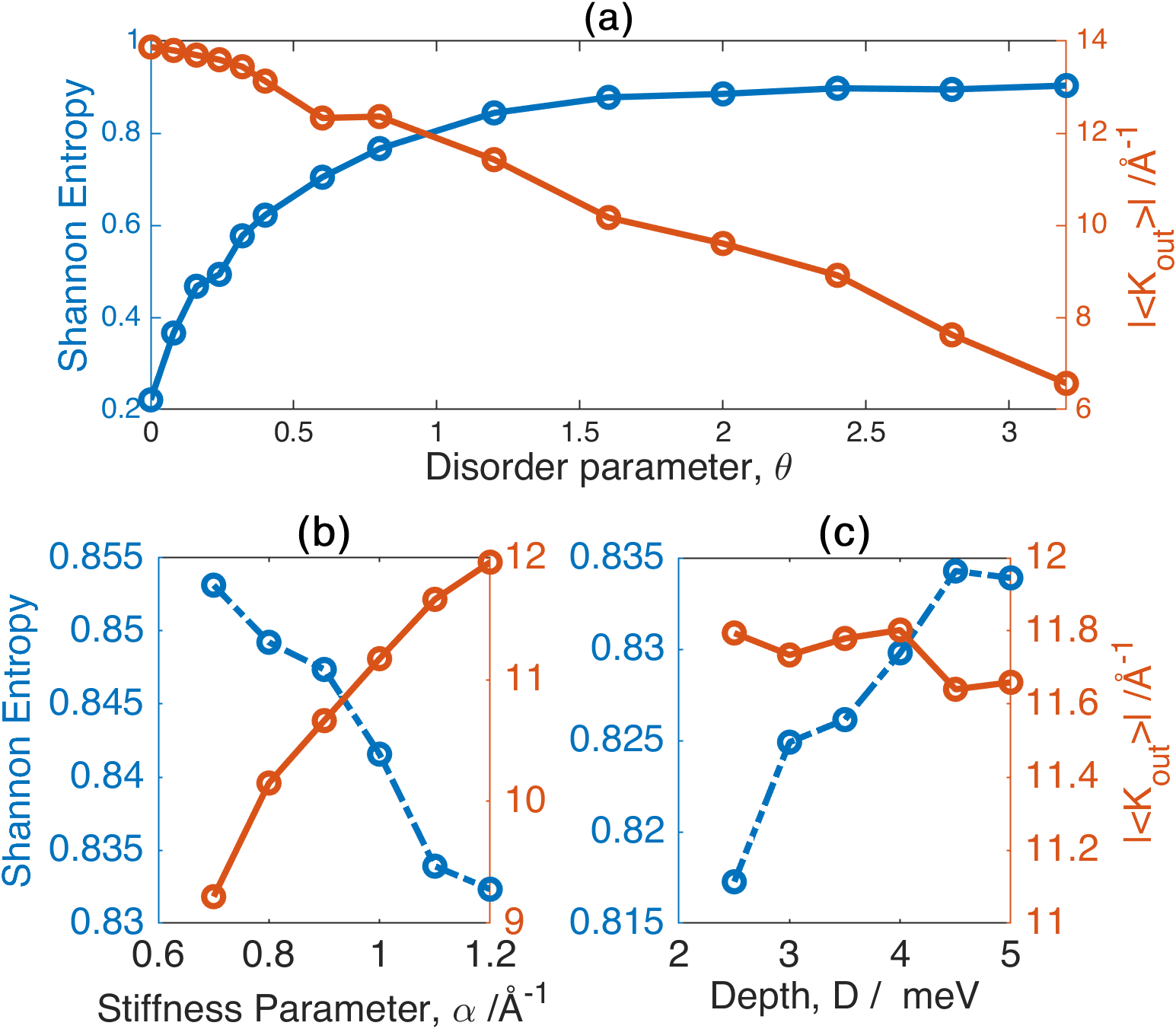}
\caption{\label{entorpy_tripple} (a) shows the observed increase in the Shannon entropy as the disorder parameter, $\theta$ was increased, as well as a decrease in \Koa. A well depth of 5meV was used and an incoming beam energy of 20meV at a $30^o$ angle to the normal scattered along the $<110>$ azimuth. (b,c) show the corresponding trends for changing the surface potential as the disorder parameter was held constant at $1.2$, whilst the stiffness parameter and well depth of the corrugated Morse potential were varied respectively.}
\end{figure}

To obtain further insight into the parameters which govern the tendency of the calculated scattering matrices towards a Knudsen scattering situation, various parameters in the He-surface interaction potential were explored. Figure \ref{entorpy_tripple} shows the change in the Shannon entropy and in the momentum average, calculated as a function of the disorder parameter, the stiffness of the potential, and the potential well-depth. Right at the start, it is clear from Figure \ref{entorpy_tripple}(a) that an increase in disorder results in an increase in the tendency towards Knudsen scattering. The Gaussians used to introduce disorder were with a height of $0.5 \angstrom$ and width of $0.5 \angstrom$. Interestingly, for these specific Gaussian parameters, the change in the Shannon entropy is very slow already after adatom concentration (disorder parameter) of about $\theta=1$, while the momentum average keeps changing in a roughly constant pace. Figure \ref{entorpy_tripple}(b) shows that the scattering tends to Knudsen the more the stiffness parameter, $\alpha$, is decreased. The smaller $\alpha$, the wider the attractive well (of the He-surface potential) in the perpendicular dimension. This result can be understood since a wider attractive well may retain the helium particle at the proximity of the surface for a longer time, allowing more time for multiple scattering in the well. Figure \ref{entorpy_tripple}(c) shows that as the depth of the attractive well is increased, the Shannon entropy increases, however, the momentum average is only weakly affected. Importantly, the spring constant of the well is also proportional to the depth $D_0$; an increase in $D_0$ results a stiffer potential, which, as shown in fig \ref{entorpy_tripple}(b) increases the momentum average, hence reducing the tendency towards Knudsen scattering. Therefore, two competing effects play a role in fig. \ref{entorpy_tripple}(c); the increase in $D_0$ and the increase in the spring constant, with the result that the momentum average is only weakly affected by $D_0$. An increase in $S$ while \Koa stays almost constant could inform on broadening the scattering probability mostly around the specular condition. It is important to note that the rate of change (of $S$ or \Koa) as a function of either $\alpha$ or $D_0$, is very small compared to the impact of additional adatoms (Gaussians).

\begin{figure}[h!]
\includegraphics[width=0.45\textwidth]{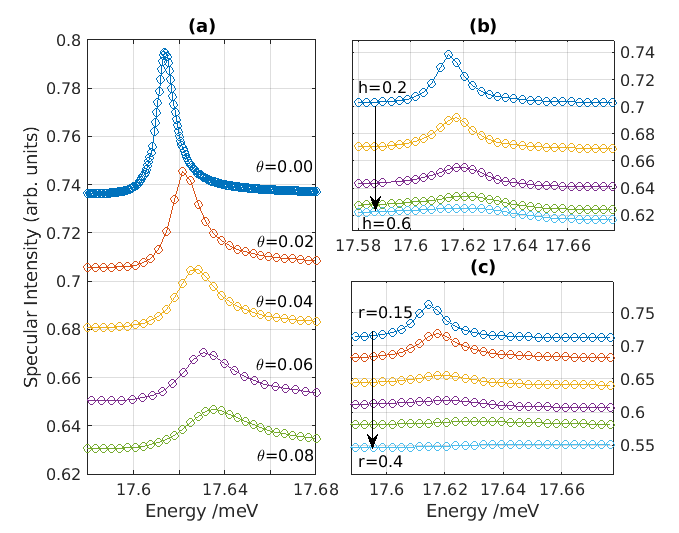}
\caption{\label{resonance_widths} (a) Energy shift in helium-surface bound-state resonances induced by surface disorder. As the disorder parameter, $\theta$ (the average number of adsorbates per unit-cell) is increased, the bound state is shifted to higher energies. (b) The effect of changing the height, h [$\angstrom$] of the Guassian-shaped adsorbates added to the surface potential, on the bound-state resonance, while the disorder parameter is fixed at 0.083. (c) The effect on the BSR of changing the radius, r [$\angstrom$] of the Guassians. For each an increase in the effective surface coverage of the disorder is seen to increase the width of the resonance, decrease the visibility as well as shift the resonance to a higher energy.}
\end{figure}

Knudsen scattering is associated with loss of memory due to multiple scattering, which requires the helium atom to spend time at the surface. In helium scattering, temporary trapping of the helium atoms is resulted due to helium-surface bound-state resonances \cite{riley2008analysis,avidor2016helium}. We have therefore explored the effect of bound-state resonances on our observation, as presented in Fig. \ref{resonance_widths}(a). For each disorder parameter, the energy of the incoming beam was varied across a bound state resonance, and the specular intensity was plotted. The blue line shows a resonance in the absence of disorder and it is seen that as the concentration of modelled adsorbates is increased, the width of the resonance increases and the peak shifts to the right. Therefore, both the width and shift in energy can inform on the nature of the disorder.

To explore the reasoning for the shift, we have explored this phenomena as function of the dimensions of the Gaussians used to introduce disorder at the surface. Figure \ref{resonance_widths}(b-c) shows the change of the BSR energy as function of the adatom width and height, and the blue-shift is clearly seen. For this particular BSR energy, the rate of blue-shift as function of adatom dimentions is similar both for the adatom width and height. Further, the rate is small compared to the change as function of the concentration of adatoms. The blue-shift phenomena can be understood since the disorder effectively narrows the spatial dimensions of the attractive well which in-turn, similar to the well known model of "particle in a box", results in an increase in the BSR energy of trapped helium atoms.

Scan curves such as the ones shown in Fig. \ref{resonance_widths} can be measured experimentally \cite{tamtogl2018helium}. Therefore, our results demonstrate the potential of using the blue-shifts and energetic width of He-surface bound-state resonances for determining the concentration of defects. In other words, this could be used for quantum metrology applications for material surfaces and 2D materials.

\begin{figure}[h!]
\includegraphics[width=0.45\textwidth]{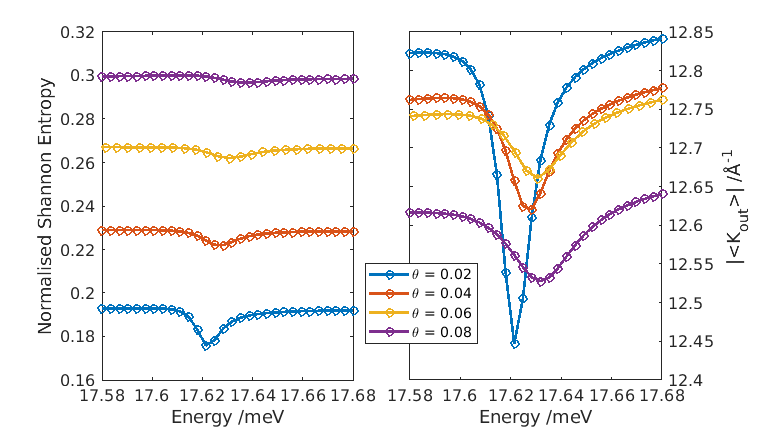}
\caption{\label{resonance_entropy_momentum} The change in the Shannon entropy $S$ (left) and in \Koa (right), as a function of the beam energy, which is swept across a bound-state resonance, for different concentrations of randomly distributed surface-defects [defects size similar to that in fig. \ref{resonance_widths}. Both $S$ and \Koa are affected by the BSR, with the effect being more significant at the low defect-concentration regime. While the trend in $S$ is expected, \Koa shows a trend opposite to expectation as discussed in the text.}
\end{figure}

We have also explored the effect of BSR on the tendency to Knudsen scattering. BSRs couple and shift scattering intensities between different diffraction channels. For example, the BSR which is explored in figure \ref{resonance_widths} enhances the specular channel (as seen in the figure) on the expense of some other diffraction channels. Potentially, some other channels are also enhanced. For this particular BSR, figure \ref{resonance_entropy_momentum} shows the change in  $S$ and \Koa as the beam energy is swept across the BSR energy. As we expect, $S$ decreases around the BSR, since the intensity is spread less evenly across the open diffraction channels. This effect means that the scattering diverge from cosine distribution, and intuitively, it is expected that \Koa will increase in such a case. However, \Koa decreases around the bound-state energy, with the decrease being strongest for a surface which is nearly defects-free. The decrease in \Koa near the BSR can be resulted by shift of the intensity to a diffraction channel which is closer to the surface normal, in addition to the enhancement of the specular channel. Overall, the QM effect of BSR is shown to significantly affect the apparent tendency towards Knudsen scattering.

Our results are relevant to the eigenstate thermalisation hypotentis (ETH). The ETH refers to the case when a quantum mechanical system can be described using equilibrium statistical mechanics \cite{Deutsch_2018}. In the current problem, we prepare a helium atom wavepacket with energy of $20 \mathrm{meV}$, far out of equilibrium. Further, we show a tendency for scattering from a disordered surface to result in distribution of the intensity, equally, to all the kinematically allowed diffraction channels (with the limit being ideal Knudsen scattering). Since all these channels correspond to the same energy, the effect can be seen as an apparent thermalisation.
Our results are obtained due to randomisation of the phase, since the helium atom scatters multiple times from the He-surface potential. Therefore, we have shown a case were the ETH can be realised without inelastic interactions, a phenomena which could be coined as an "elastic" thermalisation.

In conclusion, we have demonstrated that the tendency to Knudsen scattering can be strongly affected by quantum mechanical effects. Even more, we have shown that elastic scattering is sufficient for the phenomena to be seen, which may contribute to the ongoing discussion on the eigenstate thermalisation hypothesis (ETH). The effects are associated to the attractive part in the He-surface interaction potential, of which bound-state resonances are an important characteristic. We have demonstrated blue shifts in atom-surface bound-state energies, and pointed out to potential applications in quantum metrology of 2D materials and material surfaces.

\begin{acknowledgments}
The authors would like to thank Dr William (Bill) Allison, Prof. Ronnie Kosloff, and Dr Andrew Jardine for useful discussions. The work was conducted at the Cambridge Atom Scattering Centre (CASC), with support from the Engineering and Physical Sciences Research Council (EPSRC) via grant EP/T00634X/1. Part of this work has been performed using resources provided by the Cambridge Tier-2 system operated by the University of Cambridge Research Computing Service (www.hpc.cam.ac.uk) funded by EPSRC Tier-2 capital grant EP/P020259/1. N.A. gratefully acknowledges financial support from the Herchel Smith Fund.
\end{acknowledgments}

\bibliography{QM_Knudsen}

\end{document}